\documentclass[aps,pra,twocolumn,superscriptaddress,showpacs]{revtex4}
\usepackage{amsfonts,amssymb,amsmath}
\usepackage{graphics,graphicx,epsfig}
\usepackage{bm}
\usepackage{rotating}

\newcommand{\D}{\mathrm{d}}
\newcommand{\ds}{\displaystyle}
\newcommand{\Exp}[1]{\mathrm{e}^{\mbox{\footnotesize$#1$}}}
\newcommand{\tr}[1]{\mathrm{tr}{\left\{#1\right\}}}

\newcommand{\expect}[1]{{\left\langle{#1}\right\rangle}}
\newcommand{\KeyWords}[1]{\newline\rule{0em}{16pt}%
  {\footnotesize{Keywords:\hfill\begin{minipage}[t]{358pt}#1\end{minipage}}}}
\begin{document}
\title{Proper error bars for self-calibrating quantum tomography}
\author{Jun Yan Sim}\email[]{e0012429@u.nus.edu}
\affiliation{Centre for Quantum Technologies, %
             National University of Singapore, %
             3 Science Drive 2, Singapore 117543, Singapore}
\affiliation{Department of Physics, National University of Singapore, %
             2 Science Drive 3, Singapore 117551, Singapore}

\author{Jiangwei Shang}\email[]{jiangwei.shang@bit.edu.cn}
\affiliation{Key Laboratory of Advanced Optoelectronic  %
			 Quantum Architecture and Measurement,     %
			 Ministry of Education and School of Physics, %
             Beijing Institute of Technology, Beijing 100081, China}
\affiliation{Centre for Quantum Technologies, %
             National University of Singapore, %
             3 Science Drive 2, Singapore 117543, Singapore}
\affiliation{Naturwissenschaftlich-Technische Fakult{\"a}t, %
             Universit{\"a}t Siegen, %
             Walter-Flex-Stra{\ss}e 3, 57068 Siegen, Germany}
           
\author{Hui Khoon Ng}\email[]{cqtnhk@nus.edu.sg}           
\affiliation{Yale-NUS College, 16 College Avenue West, %
             Singapore 138527, Singapore}
\affiliation{Centre for Quantum Technologies, %
             National University of Singapore, %
             3 Science Drive 2, Singapore 117543, Singapore}
\affiliation{\mbox{MajuLab, International Joint Research Unit UMI 3654, CNRS, %
  Universit{\'e} C{\^o}te d'Azur, Sorbonne Universit{\'e}},  %
  National University of Singapore, Nanyang Technological University, Singapore}

\author{Berthold-Georg Englert}\email[]{cqtebg@nus.edu.sg}
\affiliation{Centre for Quantum Technologies, %
             National University of Singapore, %
             3 Science Drive 2, Singapore 117543, Singapore}
\affiliation{Department of Physics, National University of Singapore, %
             2 Science Drive 3, Singapore 117551, Singapore}
\affiliation{\mbox{MajuLab, International Joint Research Unit UMI 3654, CNRS, % 
  Universit{\'e} C{\^o}te d'Azur, Sorbonne Universit{\'e}}, %
  National University of Singapore, Nanyang Technological University, Singapore}
\date[]{}

%%%%%%
\begin{abstract}
Self-calibrating quantum state tomography aims at reconstructing the unknown
quantum state and certain properties of the measurement devices from the same
data. 
Since the estimates of the state and device parameters come from the same
data, one should employ a joint estimation scheme, including the construction
and reporting of joint state-device error regions to quantify uncertainty.
We explain how to do this naturally within the framework of optimal error
regions. 
As an illustrative example, we apply our procedure to the double-crosshair
measurement of the BB84 scenario in quantum cryptography and so reconstruct
the state and estimate the detection efficiencies simultaneously and reliably.
We also discuss the practical situation of a satellite-based quantum key
distribution scheme, for which self-calibration and proper treatment of the
data are necessities.
\end{abstract}

%% 03.65.Wj State reconstruction, quantum tomography
%% 02.50.-r Probability theory, stochastic processes, and statistics
%% 03.67.-a Quantum information

\maketitle

%%%%%%
\section{Introduction}\label{Sec:Intro}
Quantum tomography \cite{LNP649,Teo:16} is a basic yet crucial element in most
quantum information processing tasks.
In the typical quantum \emph{state} estimation scenario---the
focus of our paper---a finite number of quantum systems, all prepared in the
same unknown state, are probed by a measurement of one's choosing.
From the gathered data, comprising the observed sequence of measurement
detector clicks, one tries to arrive at a best guess of the unknown state.
The notion of ``best'' here depends on one's choice of figure of merit.
A popular choice is to maximize the likelihood of the data, giving the
maximum-likelihood (ML) estimator as the best guess of the state
\cite{MLEreview}.  

Standard quantum tomography strategies rely on measurement devices that are
fully characterized and well calibrated in advance, so that the unknown
quantities are solely those of the state to be estimated.
In practice, there are situations where this in-advance calibration is not
possible for all aspects of the measurement devices.
For example, the measurement device may have some properties that drift over time,
and hence require frequent recalibration.
In this case, it may be useful to do the calibration of those properties of the 
measurement device at the same time as the actual tomography of the state.
Self-calibrating quantum state tomography deals precisely with such a
situation, where the measurement devices are not fully characterized.
The properties of the measurement devices that are not calibrated in advance
and the state parameters can be reconstructed from the same gathered data.

The first initiative towards self-calibrating quantum state tomography
was taken by Mogilevtsev \textit{et al.} \cite{PRA79.020101},
with a scheme for reconstructing the quantum state and for
quantifying the mismatch between the signal and reference states. 
Later, in Ref.~\cite{PRA82.021807}, Mogilevtsev presented another scheme for
reconstructing the state and calibrating the single-photon detectors
simultaneously by exploiting some partial knowledge about the state.
By having squeezed noisy signal states in an on/off detection scheme,
this work enables absolute calibration of single-photon detectors
in a very simple way.
The first experimental realization was achieved by Bra\'{n}czyk \textit{et al.}
\cite{NJP14.085003}, who estimated the unknown rotation angle of the
measurement basis and the state parameters simultaneously.
There, the states of polarization-encoded photonic qubits were
reconstructed using wave plates with unknown retardance.
In Ref.~\cite{PRA87.062118}, Quesada \textit{et al.} showed how to circumvent
the requirement for well-characterized unitary operations in quantum state
estimation by treating unknown parameters in the state and the unitary
operations on an equal footing. 
Recently, Williams and Lougovski \cite{njp19.043003} used a Bayesian mean
estimation-based method for the simultaneous reconstruction of the unknown
state and determination of the overall detection efficiencies.  

In reporting the estimates of the various parameters, one has to include error
bars that quantify the uncertainty, and thereby make a statement about the
quality and quantity of the data, together with available prior information.
A common strategy is to deduce error bars for the measurement device
parameters and error bars (more precisely, error regions in the state space)
for the estimated state separately.
For precalibrated schemes, this is a reasonable approach as the device
parameters and the estimated state are inferred from different data; in
self-calibrating quantum tomography, however, the device and state parameters
are estimated from the \emph{same} data.
A proper reporting of uncertainty should hence involve error regions that are
regions in the combined device-state parameter space, rather than separate
error bars. 

In this paper, we discuss how to deduce proper error bars for self-calibrating
quantum tomography schemes.
The notion of optimal error regions (OERs) \cite{OER13} permits
consistent
treatment of error regions for device and state parameters within a single,
rigorous framework; since the OERs are regions around the multiparameter ML
estimator---this is an implied property, not a matter of definition---they 
naturally justify the use of ML estimators as best guesses.
The Bayesian foundation of OERs further provides the smooth incorporation of
any prior information about the measurement device and the source of the
state.
We work out examples on this matter, applying our methods, in particular, to
the scenario of satellite-based quantum key distribution (QKD), for which a
self-calibrating approach is a necessity. 
As a point of caution, we also note the additional structure in the likelihood
function, in this situation of self-calibration, that can present difficulties
for standard ML approaches.  

Below, we first begin with an overview on the self-calibration procedure, and
explain the basic notion of constructing joint ML estimators and OERs.
We then illustrate our approach in Sec.~\ref{sec:region} using the example of
the $x$-$z$ measurement scheme of the BB84 protocol for QKD \cite{BB84}.
We explore in greater detail the specific scenario of space-based QKD
experiments in Sec.~\ref{sec:qkd}, and offer concluding remarks in
Sec.~\ref{sec:conc}.

%%%%%%

\section{Basic concepts}
\subsection{Self-calibration scheme}\label{sec:scheme}
\begin{figure}[t]
\centering
\includegraphics[width=0.95\columnwidth]{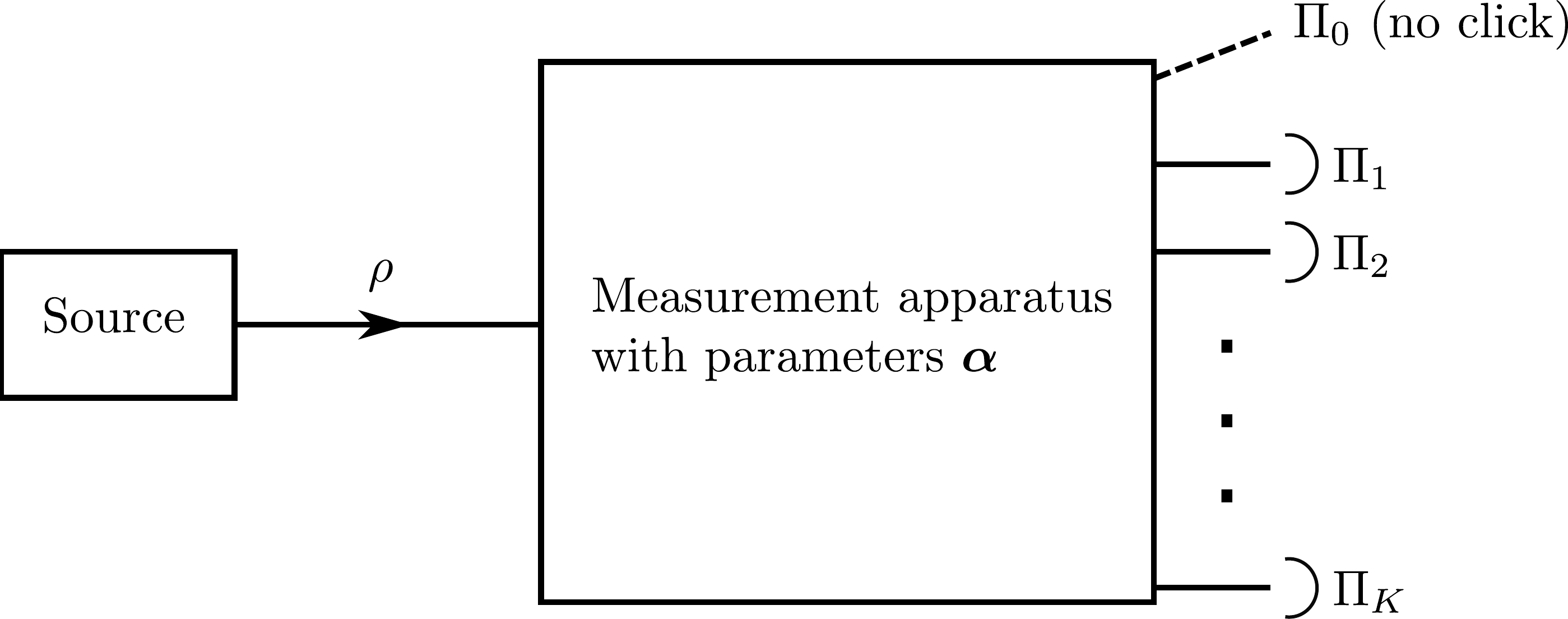}
\caption{\label{fig:TypicalScenario}%
  The typical scenario.
  The source emits identically prepared quantum systems, with their
  relevant properties described by the unknown quantum state
  $\rho$.
  A finite number of such systems are measured by the apparatus with
  probability operators $\Pi_1(\bm{\alpha})$, $\Pi_2(\bm{\alpha})$, \dots, 
  $\Pi_K(\bm{\alpha})$ for the $K$ possible outcomes.
  There is also a probability operator $\Pi_0(\bm{\alpha})$ for the null
  event. 
  The device parameters $\bm{\alpha}$ represent one's lack of knowledge about
  the apparatus.
  In a self-calibration scheme, the goal is to reconstruct $\rho$ and
  $\bm{\alpha}$ from the same data.}  
\end{figure}

In the typical scenario, as shown in Fig.~\ref{fig:TypicalScenario}, a finite
number of identically prepared quantum systems, the relevant properties of which are
described by the unknown quantum state $\rho$, are measured by the measurement
apparatus, correspondingly described by a probability-operator measurement
(POM), also known as a positive operator-valued measure.  
The POM consists of $K$ non-negative probability operators
$\Pi_1(\bm{\alpha})$, $\Pi_2(\bm{\alpha})$, \dots, $\Pi_K(\bm{\alpha})$, one
for each of the $K$ measured outcomes, and another probability operator
$\Pi_0(\bm{\alpha})$ for the null event.
The elements of the POM resolve the identity,
${\sum_{k=0}^{K} \Pi_k(\bm{\alpha}) =1}$.
The unknown parameters $\bm{\alpha}$ represent one's lack of knowledge
about the measurement devices.
Then, the probability for the $k$th outcome to occur is given by the Born rule,
\begin{equation} \label{eq: born}
  p_k = \tr{\Pi_k(\bm{\alpha})\rho} = \expect{\Pi_k(\bm{\alpha})}\,.
\end{equation}
The data $D$ are made up of the sequence of counts of the detection events,
${D=\{n_1,n_2,\dots,n_{K}\}}$, with ${N = \sum_{k=1}^K n_k}$ total counts of
detection events. 
In scenarios with imperfect detectors, there are also missing counts denoted
by $n_0$.
If we know the total number of copies sent into the apparatus, then $n_0$ is
known once we know $N$; more typically, one does not have precise information
about the total number of copies, and $n_0$ is an unknown parameter that
enters the model for the experiment. 

In quantum state estimation, $\bm{\alpha}$ is precalibrated and the task is
to reconstruct $\rho$ from the data.   
In quantum parameter estimation, $\rho$ is known and the task is to
reconstruct $\bm{\alpha}$ from the data.  
In a self-calibration scheme, one aims at reconstructing the state $\rho$ and
the device parameters $\bm{\alpha}$ from the same data.

\subsection{Maximum-likelihood estimation}
The ML estimators for the state and device parameters---denoted by
$\hat{\rho}_{\text{ML}}$ and $\hat{\bm{\alpha}}_{\text{ML}}$,
respectively---can be obtained by maximizing the likelihood function over the
joint state and device parameter space using an iterative algorithm
\cite{MLEreview}.  
For the scenarios with missing counts, the likelihood of obtaining the
observed data $D$, given the state $\rho$ and the device parameters
$\bm{\alpha}$, is 
\begin{equation}
L(D|\rho,\bm{\alpha}) = \sum_{n_0=0}^{\infty} L(D,n_0|\rho,\bm{\alpha})\,,
\end{equation}
where $L(D,n_0|\rho,\bm{\alpha})$ is the corresponding likelihood of obtaining
the data $D$ and also having $n_0$ null events.

For our present discussion of general concepts and methods, we are   content
with the scenario sketched in Fig.~\ref{fig:TypicalScenario}. 
We note, however, that there can also be unknown parameters of the source that
enter the likelihood function but are not accounted for by the quantum state
$\rho$, and source parameters of this kind can be handled analogously to the
apparatus parameters $\bm{\alpha}$. 
An example is provided by the situation discussed in Sec.~\ref{sec:qkd}.

\subsection{Optimal error regions}
The ML estimators are point estimators which represent our best guess for the
unknown quantum state and device parameters.
The point estimator calculated from a finite amount of data
will not coincide exactly with the true parameters.
To be statistically meaningful, a point estimator should be endowed
with error bars---error regions in higher dimensions---and these regions
should be optimally chosen by appropriate criteria.
In the Bayesian approach used here, the OERs can be equivalently specified
as having either largest credibility ($\equiv$ posterior content) for the
prechosen size ($\equiv$ prior content) or smallest size for the prechosen
credibility; see Refs.~\cite{Evans+2:06,OER13}.
The OERs can be characterized very easily, as they are bounded-likelihood
regions---regions containing all the points with their likelihood larger than
or equal to some threshold value, 
\begin{equation}\label{eq:BLR}
  \mathcal{R}_\lambda(D)
  =\{(\rho,\bm{\alpha})\,|\,L(D|\rho,\bm{\alpha})\ge
  \lambda L_{\text{max}}(D)\}
\end{equation}
with $0\le\lambda\le1$, where $L_{\text{max}}(D)%
=L(D|\hat{\rho}_\text{ML},\hat{\bm{\alpha}}_\text{ML})$
is the maximum value of the likelihood function.
$\mathcal{R}_0$ is the same for all data $D$; it
contains all thinkable quantum states and device parameter values.
Owing to the simple inequality in Eq.~(\ref{eq:BLR}), it is easy to
check whether a particular $(\rho,\bm{\alpha})$ of interest is inside the
$\mathcal{R}_\lambda$ in question, even if the high-dimensional OERs are
difficult to visualize.

For any OER $\mathcal{R}_\lambda$, there are two important quantities, its
size $s_\lambda$ and its credibility $c_\lambda$. 
The size of a region is the assigned probability of finding the true state and
device parameters in the region, prior to  acquiring the data.
For the OERs, then,
\begin{equation}\label{eq:s-lambda}
  s_\lambda(D)= \int\limits_{\mathcal{R}_\lambda(D)}
  (\mathrm{d}\rho)(\mathrm{d}\bm{\alpha})
\end{equation}
with
\begin{eqnarray}
  (\mathrm{d}\rho)&=&\mathrm{d}q_1\cdots\mathrm{d}q_J\,
  w_\text{cstr}(q_1,\dots,q_J)w_0(q_1,\dots,q_J)\,,
\nonumber\\
(\mathrm{d}\bm{\alpha})&=&\mathrm{d}\alpha_1\cdots\mathrm{d}\alpha_M
 w_0(\alpha_1,\dots,\alpha_M)\,,
\end{eqnarray}
where $q_1,\dots,q_J$ denote the state parameters;
the factor $w_\text{cstr}(q_1,\dots,q_J)$ accounts for all the constraints
that the state parameters have to satisfy;
$w_0(q_1,\dots,q_J)$ and $w_0(\alpha_1,\dots,\alpha_M)$ are the
prior densities which represent our knowledge on the state and device
parameters.

After the data have been obtained, we update our belief about the state and
device parameters by multiplying the prior density with the likelihood
function to obtain the posterior density. 
The credibility of a region is the probability of finding the true state and
device parameters in the region, conditioned on the observed data.
For the OERs, then, 
\begin{equation}\label{eq:c-lambda}
  c_\lambda(D) = \frac{1}{L(D)}\int\limits_{\mathcal{R}_\lambda(D)}
  (\mathrm{d}\rho)(\mathrm{d}\bm{\alpha})\, L(D|\rho,\bm{\alpha})\,,
\end{equation}
where
\begin{equation}\label{eq:LD}
L(D) = \int_{\mathcal{R}_0}(\mathrm{d}\rho)(\mathrm{d}\bm{\alpha})\,
L(D|\rho,\bm{\alpha})  
\end{equation}
is the likelihood of the data $D$.
We have ${c_\lambda\ge s_\lambda}$, with the equal sign usually only holding
for ${\lambda=0}$ and $1$, and there is the link
\begin{equation}
 c_{\lambda}=\frac{\ds\lambda s_{\lambda}+\int_\lambda^1 \D \lambda'
 \,s_{\lambda'}} {\ds\int_0^1 \D \lambda' \,s_{\lambda'}}
\end{equation}
between $c_{\lambda}$ and $s_{\lambda}$,
so that sometimes only the first of the high-dimensional integrals in
Eqs.~(\ref{eq:s-lambda}), (\ref{eq:c-lambda}), and (\ref{eq:LD}) needs to be
computed numerically.

A related concept is the plausible region \cite{Evans:15, OEI16} which is
based on the principle of evidence.
The principle of evidence states that if the posterior probability of a
$(\rho,\bm{\alpha})$ pair is larger than its prior probability then the data
give evidence in favor of the $(\rho,\bm{\alpha})$ pair,
and the data give evidence against the pair if the posterior probability 
is less than the prior probability.
The plausible region is composed of all the $(\rho,\bm{\alpha})$
pairs, in favor of which the data give evidence---all the pairs with
$L(D|\rho,\bm{\alpha})>L(D)$. 
Accordingly, the plausible region is the OER for the critical $\lambda$ value 
\begin{equation}
  \lambda_\text{crit}(D)\equiv\frac{L(D)}{L_\text{max}(D)}
  =\int_0^1\D\lambda\, s_{\lambda}\,.
\end{equation}
Once a prior has been chosen, the plausible region is solely determined
by the data.
As more and more data are obtained, the size of the plausible region
decreases while its credibility increases. 
If the size $s_{\lambda_{\text{crit}}}$ of the plausible region is small and
its credibility $c_{\lambda_{\text{crit}}}$ is large, ``then this suggests an
accurate estimate has been obtained'' \cite{Evans-essay-2019}.
One can quantify the strength of the evidence associated with an OER
in terms of a suitably chosen function of the ratio
$c_{\lambda}/s_{\lambda}$; see Ref.~\cite{Evans-Guo-2019} for a discussion
of various aspects thereof. 

To report the error region for the observed data $D$, following the scheme in
Ref.~\cite{OER13}, $s_\lambda$ and $c_\lambda$ are plotted as functions of
$\lambda$; see Figs.~\ref{fig:normals&c} and \ref{fig:extremes&c} below.
For any desired level of credibility, the corresponding $\lambda$ and the size
of the OER can be determined from the graph.
Alternatively, if the experimenter chooses to report the plausible region, the
critical $\lambda$ value can be calculated and the size and credibility of the
plausible region can be read off from the graph.
The difference ${c_\lambda-s_\lambda}$ is largest for
${\lambda=\lambda_\text{crit}(D)}$.

To calculate the size and credibility of a region for the observed data $D$,
one has to resort to Monte Carlo integration (see Refs.~\cite{Shang+1:15,
  Seah+1:15, YS19} for the application of Monte Carlo integration to this
context of OERs). 
Random samples in the joint state and device parameters space have to be
generated to perform the high-dimensional integrals for the size and
credibility.
We do this by employing the Hamiltonian Monte Carlo strategy described
in Ref.~\cite{Seah+1:15}.

\begin{figure*}
  \centering
  \begin{tabular}{@{}r@{\qquad}r@{\ }l@{}}
    \raisebox{-20pt}{\includegraphics[width=0.65\textwidth]{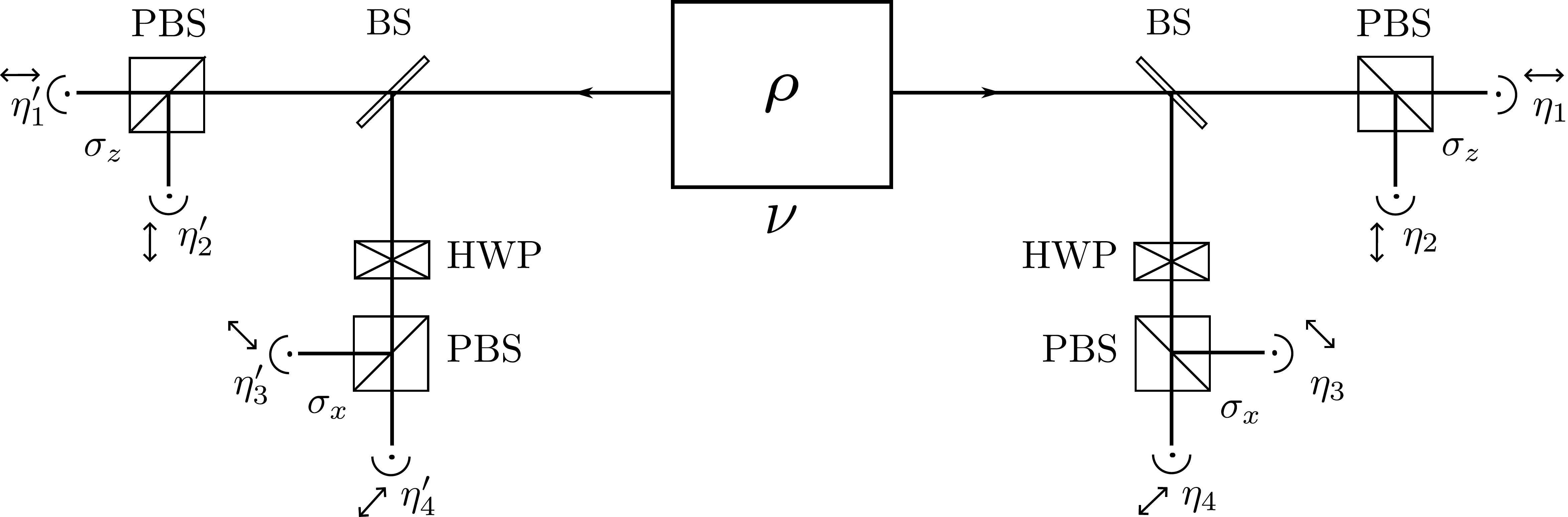}}&
   \raisebox{5pt}{\begin{turn}{90}%
                  \parbox{5.2em}{detection on the left}\end{turn}}&
 $\ds\begin{array}[b]{c|ccccc}
   \multicolumn{6}{r}{\text{detection on the right}}\\[0.5ex]
    & 1 & 2 & 3 & 4 & \text{null}\\\hline
  1'& n_1 & n_2 & n_3 & n_4 & n_5\\
  2'& n_6 & n_7 & n_8 & n_9 & n_{10} \\
  3'& n_{11} & n_{12} & n_{13} & n_{14} & n_{15} \\
  4'& n_{16} & n_{17} & n_{18} & n_{19} & n_{20} \\
  \text{null}& n_{21} & n_{22} & n_{23} & n_{24} & 
\end{array}$
\end{tabular}
\caption{\label{fig:BB84}
  Setup for the double-crosshair measurement of the BB84 scheme.
  A photon source emits polarization-entangled photon pairs that follow a Poissonian
  distribution with mean number $\nu$.
  The photons are sorted by polarizing beam splitters (PBSs) and
  then detected; on the way to the respective PBS, each photon has an equal
  chance of being reflected or transmitted by a beam splitter (BS) and, when
  reflected, has its polarization turned by a half-wave plate (HWP).
  Accordingly, each of the photons is eventually measured in either the
  $\sigma_x$ or the $\sigma_z$ basis (visualized as a crosshair in the $x$-$z$
  plane of the Bloch ball) with equal probability.  
  The eight detectors have nonunit efficiencies denoted by $\eta$, so that
  some photons escape detection.
  The table on the right shows how the event counts $n_1$, $n_2$, \dots,
  $n_{24}$ refer to coincidence events or single detection events.
  There is no 25th entry because the number of double-null events is not known.
}
\end{figure*}

\section{Self-calibration for a BB84 experiment}\label{sec:region}
To explain our approach, and illustrate the significance of proper error
regions in self-calibration schemes, we consider a concrete example: a BB84
QKD experiment \cite{BB84}.
We are envisioning an experimenter wishing to set up a BB84 QKD protocol, who
needs to calibrate the entangled photon source, as well as the detectors to be
used in the protocol.
Such knowledge is needed for protocols like BB84 that do not have the
device-independence properties of more recent QKD schemes; yet, even
experimenters setting up device-independent schemes would potentially require
such self-calibration experiments to understand their own setup, even if the
information is not used in the security analysis.
We first set up the problem, and discuss the general approach to constructing
proper error regions; it should be emphasized that, while we discuss our
approach for this specific BB84 example, our methods apply to other
situations. We discuss, as a final subsection, a practical application to
space-based QKD, where such self-calibration is a necessity.

\subsection{Setup}
In one version of the BB84 scheme, a photon source emits entangled photon
pairs that follow a Poissonian distribution with mean number $\nu$; see 
Fig.~\ref{fig:BB84}.
One of the photons in each pair is sent to a measurement apparatus on the
left; the other photon is sent to the apparatus on the right.
On each side, the photon is measured by a four-outcome ``crosshair'' POM
with the ideal detection probabilities
\begin{eqnarray}\label{eq:A1}
\left.\begin{array}{l}
p_{1}\\
p_{2}
\end{array}\right\} = \frac{1}{4}(1\pm z)\,,\quad
\left.\begin{array}{l}
p_{3}\\
p_{4}
\end{array}\right\} = \frac{1}{4}(1\pm x)\,,
\end{eqnarray}
where $x=\expect{\sigma_x}$ and $z=\expect{\sigma_z}$ are the expectation values
of two components of the Pauli vector operator $\bm{\sigma}$.
These expressions would apply to the photon detection on the right in
Fig.~\ref{fig:BB84} if the detectors had unit efficiency.
When accounting for the finite detection efficiencies, we have expressions
such as
\begin{equation}\label{eq:A2}
  \eta_2\eta'_{3}\expect{\frac{1-\sigma_z}{2}
                                     \otimes\frac{1+\sigma_x}{2}}
\end{equation}
for the probability of detecting the photons of a pair by detectors $2$ on the
right and $3'$ on the left in coincidence.
Here, $\eta_k$ is the probability that detector $k$ functions correctly, i.e.,
detects a photon that falls on it.
For simplicity, we are here assuming that there is a negligible chance of
losing the photon on the way from the source to the detector; this is
markedly different in the situation of Sec.~\ref{sec:qkd}.

Owing to the imperfection of the detectors, there are actually five possible
outcomes on each side including a null event, thus 25 different joint
outcomes in total. 
However, the double-null events where both photons escape detection are not
recorded and, since the actual number of entangled photon pairs is not known,
we also do not know how many double-null events have occurred.
Therefore, the data $D$ are made up of the sequence of 24 counts of
detection events, $D=\{n_1,n_2,\dots,n_{24}\}$; see the table in
Fig.~\ref{fig:BB84}.
In this scenario, the likelihood for obtaining the data $D$, 
given the state $\rho$ and detector efficiencies $\bm{\eta}$, is
\begin{equation}\label{eq:likelihood}
  L(D|\rho,\bm{\eta}) = \Exp{\nu p_0}\prod\limits_{k=1}^{24}{p_k}^{n_k}\,,
\end{equation}
up to an overall factor of no consequence, where $p_k$ is the probability of
detecting an event of the $k$th kind and
\begin{equation}
  \label{eq:B2}
  p_0=1-\sum_{k=1}^{24}p_k
\end{equation}
is the probability of getting a double-null event; 
see the Appendix for a derivation.

In this example, the only unknown parameters of the measurement devices are
the detector efficiencies; other aspects of the measurement devices are
assumed to have been precalibrated. 
So, our task is to infer the eight state parameters---$\expect{1\otimes\sigma_x}$,
$\expect{1\otimes\sigma_z}$, $\expect{\sigma_x\otimes 1}$,
$\expect{\sigma_z\otimes 1}$, $\expect{\sigma_x\otimes \sigma_x}$,
$\expect{\sigma_x\otimes \sigma_z}$, $\expect{\sigma_z\otimes\sigma_x}$,
and $\expect{\sigma_z\otimes\sigma_z}$---and the eight detector
efficiencies---$\eta_1$, $\eta_2$, $\eta_3$, $\eta_4$, $\eta'_1$, $\eta'_2$,
$\eta'_3$, and $\eta'_4$---from the data, and not only report our best guess
for these 16 numbers but also quantify the accuracy of the inferred
values.
As it turns out, estimating all eight detector efficiencies using the ML
approach cannot be done in a straightforward manner: The resulting likelihood
function can have multiple maxima, which renders ML estimation ambiguous.
We elaborate on this matter in the next section.  

To focus our discussion on the issue of proper error regions, rather than
resolving this ambiguity in the ML estimation scheme (worthy of further
investigation elsewhere), we instead assume a simplification in the form of
prior information about the detectors.
Specifically, we assume that the detector efficiencies on each side of the
setup are in stable, and precalibrated, ratios with one another.
In Sec.~\ref{sec:qkd}, we describe a physical scenario for which such prior
information about the detectors is natural.
Then, what is unknown and to be estimated in the self-calibration scheme, are
the maximum detector efficiencies, one for each side,
$\eta_\textrm{right}=\max\{\eta_1,\eta_2,\eta_3,\eta_4\}$ and
$\eta_\textrm{left}=\max\{\eta'_1,\eta'_2,\eta'_3,\eta'_4\}$.
Once these efficiencies are known, we know the individual detector
efficiencies from our precalibrated ratios.

%%%%%%
\subsection{Multiple local maxima of the likelihood function}\label{sec:local maxima}
\begin{figure}[b]
\centering
\includegraphics[width=0.95\columnwidth]{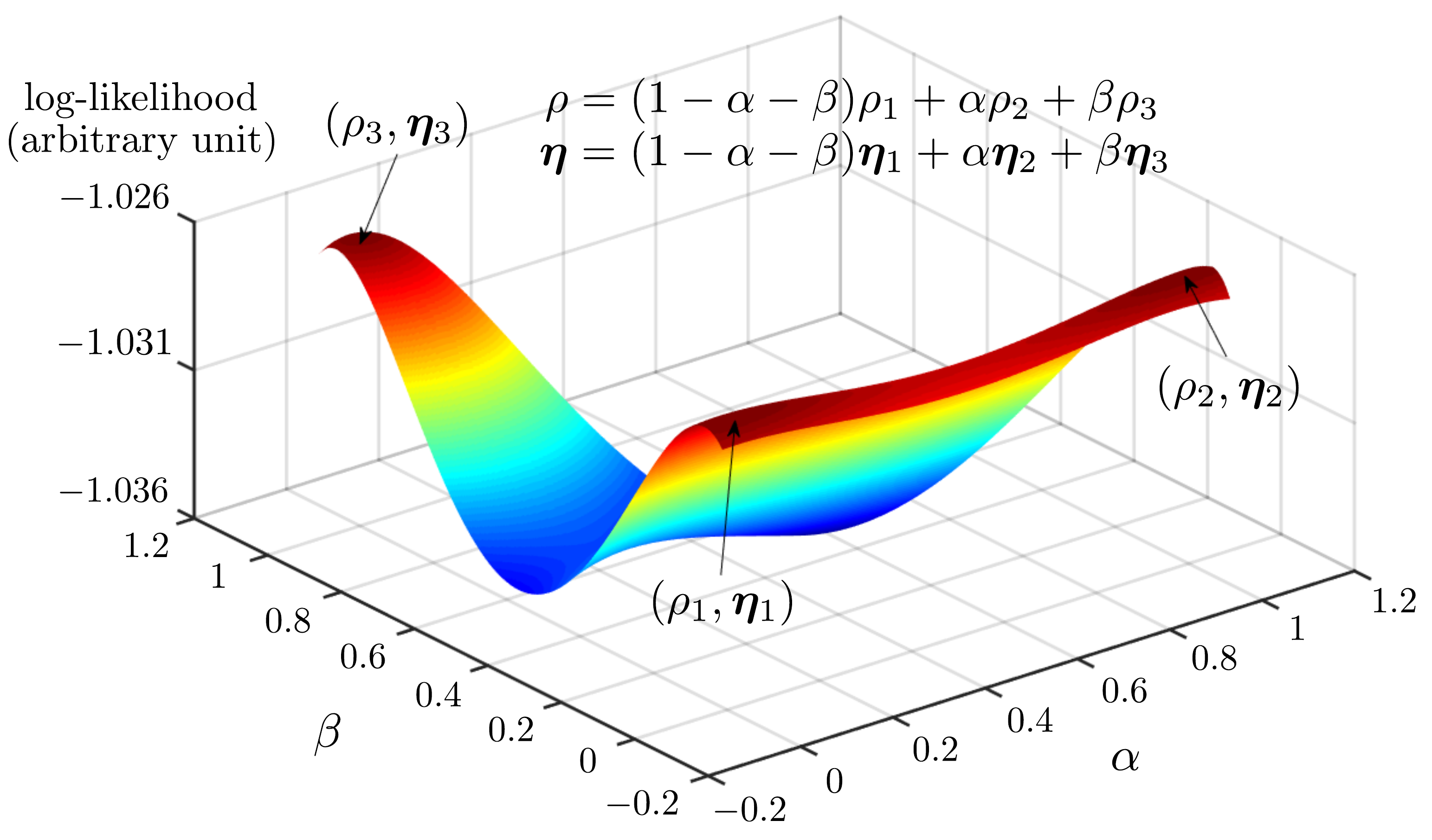}
\caption{\label{fig:maxima}%
An instance of having multiple maxima in the likelihood function, where
$(\rho_1,\bm{\eta}_1)$, $(\rho_2,\bm{\eta}_2)$, and $(\rho_3,\bm{\eta}_3)$
identify the three maxima.
Log-likelihood values of the convex combination of the three maxima are
plotted. It is clear from the graph that the log-likelihood function is not
concave and has several maxima.
Note that the probabilities that enter the likelihood in
Eq.~(\ref{eq:likelihood}) are different for the three maxima; the occurrence
of multiple maxima does not result from an ambiguity in mapping the
probabilities on $\rho$ and $\bm{\eta}$.}
\end{figure}

When we do not take the prior information on the ratios of detector
efficiencies into account and try to reconstruct the eight efficiencies and
the eight state parameters with ML estimation, we observe multiple maxima in
the likelihood function in some cases, which are neither typical nor rare. 
These maxima have approximately the same height, but refer to
very different situations as they are usually far from
each other in the joint parameter space.
We illustrate this feature with an example from a simulated experiment,
see Fig.~\ref{fig:maxima}.
The data obtained in the simulation are
$D=\{$597, 515, 658, 345, 2012, 1039, 804, 1081, 506, 3308, 1091, 795, 990,
600, 3446, 740, 558, 826, 326, 2334, 4228, 3123, 4295, 2143$\}$. 
As can be seen, the three maxima are far away from each other.
In particular, if we choose two of them, the fidelity between the two states
is not high,
${F(\rho_1,\rho_3)=\tr{{\left|\sqrt{\rho_1}\sqrt{\rho_2}\right|}}=0.84}$,
and the efficiencies are rather distinct, $|\bm{\eta}_1-\bm{\eta}_3|= 0.97$. 

The presence of multiple maxima indicates that the log-likelihood function
is not concave in the joint space of quantum states and efficiencies.
A unique ``best guess'' based on maximizing the likelihood is then not
available, even if there is---mathematically speaking---a single global
maximum because local maxima may have parameter values that are equally
plausible for a physicist.
This is the situation illustrated by the example in Fig.~\ref{fig:maxima}, where
the self-calibrating scheme for the double-crosshair measurement
does not yield an unambiguous answer, and the OERs $\mathcal{R}_\lambda$
consist of disjoint pieces when $\lambda\lesssim1$; the plausible region
is also of this kind. 
While it is true that the global maximum tends to dominate when 
sufficiently many events are observed, and then the resulting plausible
region is convex, the situation can easily be inconclusive for a large, but
not that large, number of observations.

Yet, an ambiguity of this kind can often be resolved by taking additional
information into account.
Rather than using it for choosing among the candidate estimators identified
by maximizing the likelihood, it is preferable, if possible, to reduce the
number of parameters.
As explained earlier, in the example studied here, this reduction is achieved
by prior knowledge about the ratios of the four detection efficiencies on each
side of the scheme in Fig.~\ref{fig:BB84}.

Ratios of the detector efficiencies can be determined much more easily than
the absolute values.  
For example, by counting photons from the same source for the same period
of time by each of the detectors, the ratios of their efficiencies
are simply the ratios of the photon counts.
Therefore, we modify our scheme such that ratios of the detector efficiencies
on each side of the setup are measured prior to the BB84 experiment.

When the prior information of knowing ratios of the detector efficiencies is
included, we are left with the eight state parameters, plus two efficiencies,
i.e., $\eta_\text{right}=\max\{\eta_1,\eta_2,\eta_3,\eta_4\}$
and $\eta_\text{left}=\max\{\eta'_1,\eta'_2,\eta'_3,\eta'_4\}$ to be estimated
from the data.
All the other detector efficiencies can then be deduced from the estimates for
$\eta_\text{right}$ and $\eta_\text{left}$.

Simulations provide strong numerical evidence that the likelihood function for
the new scheme has only one maximum; it is log-concave indeed. 
For the example in Fig.~\ref{fig:maxima}, a unique estimator
$(\hat{\rho}_\text{ML}$, $\hat{\bm{\eta}}_\text{ML})$ is found after the prior
knowledge  
of ratios of the detector efficiencies is taken into account.
Moreover, this estimator is very close to the mock-true state $\rho_\text{true}$
and the efficiencies $\bm{\eta}_\text{true}$ used for generating the simulated data.
Specifically, we have the fidelity $F(\hat{\rho}_\text{ML},\rho_\text{true})=0.9999$,
and $|\hat{\bm{\eta}}_\text{ML}-\bm{\eta}_\text{true}|=0.0019$.

For the examples in this paper, we assume the specific values of ratios of
detector efficiencies can be obtained from precalibration measurements.
In the more general situation, rather than prescribing specific values for the
ratios of detector efficiencies, one can also remove the multiple local maxima
from the problem by prescribing a \emph{distribution} for the ratios.
We first reparametrize $(\rho,\bm{\eta})$ as $(\bm{s},\bm{k})$, where
$s=(\rho,\eta_\text{left},\eta_\text{right})$ as before, and $\bm{k}$ are the
ratios of detector efficiencies.
We then regard $\bm{k}$ as nuisance parameters to be integrated out, according
to a prescribed distribution deduced from precalibration measurements, so
that we are again left with only $\bm{s}$ to estimate from our data.  
Our numerical investigations suggest that, for chosen fixed values of
$\bm{k}$, the likelihood function $L(D|\bm{s},\bm{k})$ is concave in the space
of $\bm{s}$, 
\begin{align}
  L(D|\lambda\bm{s}_1+(1-\lambda)\bm{s}_2,\bm{k})\geq
  &\lambda  L(D|\bm{s}_1,\bm{k})  \\\nonumber
&\mbox{}+ (1-\lambda)L(D|\bm{s}_2,\bm{k}),
\end{align} 
for $\lambda\in[0,1]$ and $\bm{s}_1$ and $\bm{s}_2$ are two values of $\bm{s}$. 
Integrating over the ratios $\bm{k}$ according to a prior distribution
$(\mathrm{d} \bm{k})w(\bm{k})$ gives the marginal likelihood,  
$L(D|\bm{s}) \equiv \int (\mathrm{d}\bm{k}) w(\bm{k}) L(D|\bm{s},\bm{k})$,
which, by linearity, inherits this concavity property and has a single maximal
point.

%%%%%%

\subsection{ML estimation and proper error regions}
\begin{figure}[t]
\centering
\includegraphics[width=0.95\columnwidth]{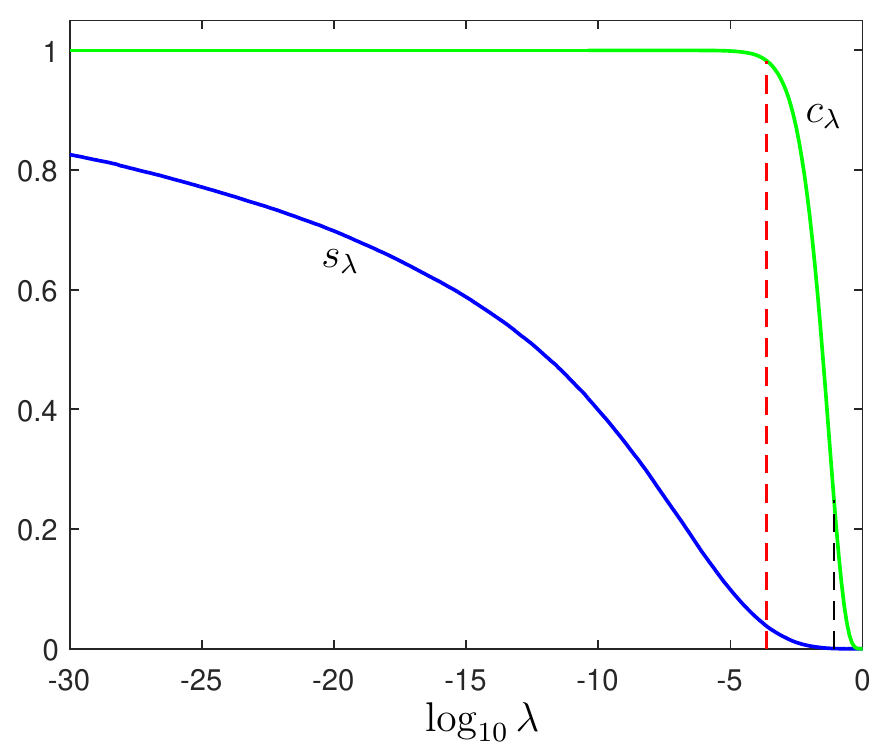}
\caption{\label{fig:normals&c}
Size $s_\lambda$ and credibility $c_\lambda$ of the ten-dimensional
OERs $\mathcal{R}_\lambda$ as a function of $\log_{10}\lambda$.
The red vertical dashed line marks the critical $\lambda$ value at
$\lambda_\text{crit}=2.34\times 10^{-4}$, which identifies the plausible region.
The black vertical dashed line marks the $\lambda$ value for the mock-true state
and efficiencies at $\lambda_\text{true}=8.27\times 10^{-2}$.}
\end{figure}
\begin{table}[b]
\centering
\caption{Mock-true and ML-estimated values of the 10 parameters used for the
  simulated data of Fig.~\ref{fig:normals&c}. Since there are only 66
  detection events in this simulation, the estimated values do not agree well
  with the mock-true values.} 
\label{tab:mle1}
\begin{tabular}{c @{\qquad} c @{\qquad} c}
  Parameter & Mock-true value
            & ML-estimated value   \\ \hline
$\expect{1\otimes\sigma_x}$ & -0.1201 & -0.2658 \\
$\expect{1\otimes\sigma_z}$ & -0.0803 & -0.0578\\
$\expect{\sigma_x\otimes 1}$ & -0.0592 & \ 0.2200 \\
$\expect{\sigma_z\otimes 1}$ & \ 0.3783 & \ 0.1643 \\
$\expect{\sigma_x\otimes \sigma_x}$ & -0.0182 & -0.0736 \\
$\expect{\sigma_x\otimes \sigma_z}$ & \ 0.4009 & \ 0.5693 \\ 
$\expect{\sigma_z\otimes\sigma_x}$ & -0.0434 & \ 0.0488 \\
$\expect{\sigma_z\otimes\sigma_z}$ & \ 0.1359 & -0.1060 \\
$\eta_\text{left}$ & \ 0.6755 & \ 0.5831\\
$\eta_\text{right}$ & \ 0.7746 & \ 0.6565
\end{tabular}
\end{table}

In one simulated experiment for the double-crosshair measurement of the BB84
scheme, we obtained the data 
$D=\{$1, 2, 1, 1, 7, 1, 1, 0, 0, 2, 3, 3, 1, 1, 7, 1, 0, 1, 3, 1, 5, 12, 3, 9$\}$. 
In this example, we assume that we do not know anything about the mock-true state, 
$\eta_\text{left}$, and $\eta_\text{right}$ before the data were taken. 
Thus, the prior we choose is uniform in the eight state parameters, and
also uniform in $\eta_\text{left}$ and $\eta_\text{right}$.
Figure~\ref{fig:normals&c} shows the plot for the size and credibility of the
ten-dimensional OERs $\mathcal{R}_\lambda$ as a function of
$\log_{10}\lambda$, calculated by a Monte Carlo integration that uses a random
sample with $500\,000$ points.
To ensure the physicality of the states in the Hamiltonian Monte Carlo
sampling, we use the parametrization described in Sec.~4.3 of
Ref.~\cite{Seah+1:15}.  
The size and credibility of the plausible region are ${s=0.0378}$ and 
${c=0.9826}$, respectively. 
The mock-true state and efficiencies used for the simulation are contained in
the OERs with ${\lambda<8.27\times 10^{-2}}$ and ${c>0.249}$.  
Thus, they are in the plausible region.

More specifically, this simulation used detector efficiencies with the
following ratios: 
\begin{equation}
\eta'_{1}:\eta'_{2}:\eta'_{3}:\eta'_{4} = 0.4172 : 0.5510 : 1 : 0.6777\,,
\end{equation}
and
\begin{equation}
\eta_{1}:\eta_{2}:\eta_{3}:\eta_{4} = 0.6595 : 1 : 0.6287 : 0.7619\,,
\end{equation}
and the parameters of the mock-true state are reported in the middle column of
Table~\ref{tab:mle1}.
This table also shows the parameter values of the ML estimators
$\hat{\rho}_\text{ML}$ and $\hat{\eta}_\text{left}$,
$\hat{\eta}_\text{right}$. 
We note that, although the estimators are in the plausible region, the
estimated parameter values are rather different from the ones used for the
simulation.  
This is not unexpected for so few data, namely, only 66 detection events.

\begin{figure}[t]
\centering
\includegraphics[width=0.95\columnwidth]{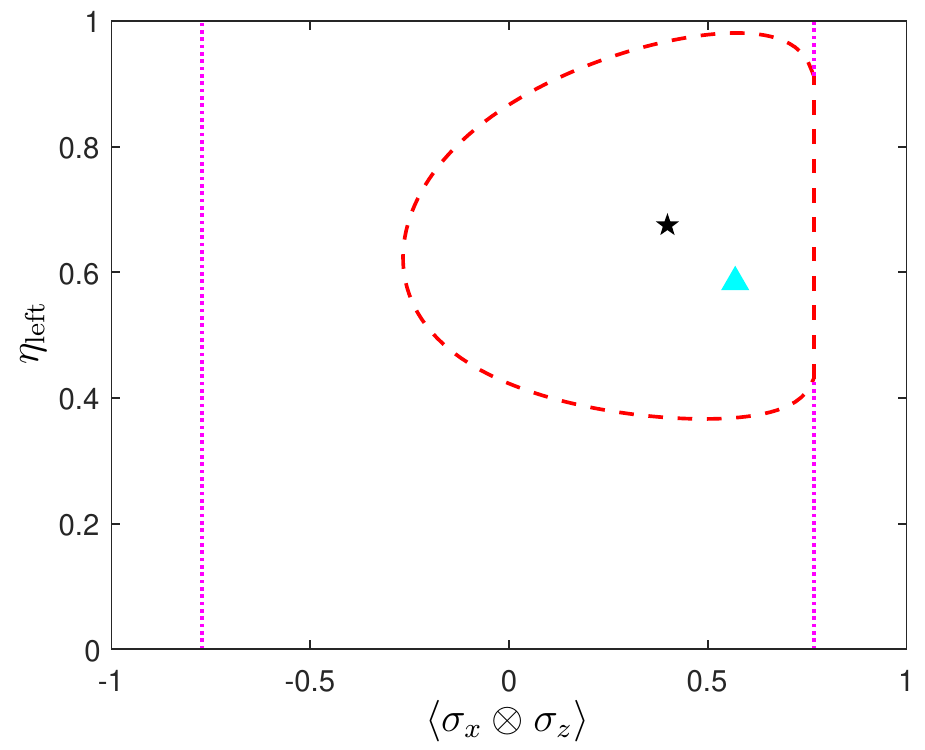}
\caption{\label{fig:2dRegion}
The dashed red contour shows the boundary of the
two-dimensional slice of the plausible region.
The other parameters are set to their mock-true values.
The black star and the cyan triangle show
the mock-true values and ML estimates, respectively.
The permitted values for $\expect{\sigma_x\otimes\sigma_z}$ are those
between the dashed vertical magenta lines; it is not the full range from
$-1$ to $+1$ because of the restrictions imposed by the mock-true values of
the other expectation values.}
\end{figure}

As an illustration that the error regions are regions in the joint
device-state parameter space, Fig.~\ref{fig:2dRegion} shows the
two-dimensional slice of the ten-dimensional plausible region with the other
parameters set to their mock-true values.
Observe that there is no natural way to break up the error region into a
device-parameter-only region, and a state-parameter-only region.
A naive approach might be to report the maximum width of a
OER of a chosen credibility or plausible region along the
$\eta_{\textrm{left}}$ axis as the error bar for that device quantity, and
then construct the state-only error region as usual. 
However, this is not representative of the actual situation where both state
and device parameters are inferred from the same data.

The correct procedure for determining an optimal error range for a single
parameter is based on the likelihood of observing the data, conditioned solely
on the parameter of interest. 
This inferred likelihood is obtained from $L(D|\rho,\bm{\alpha})$ by
marginalizing all other parameters; see Ref.~\cite{OEI16}. 
We shall not elaborate on this matter here, except for noting that, usually,
the best guess for a singled-out parameter is not equal to the best guess for
this parameter when it is estimated together with all others.

%%%%%%
\subsection{Application to space-based QKD experiments}\label{sec:qkd} 
We turn to a practical scenario where self-calibration is crucial for the QKD
experiment to function properly: space-based QKD. 
One of the main challenges in QKD is to extend the coverage towards a global
scale. 
Optical fibers and free-space links between ground stations have distance
limits due to losses in fibers or the need of line-of-sight locations.
To establish a global quantum communication network, the usage of satellites
as transmitters and receivers has been proposed \cite{space}, and there have
been successful experiments demonstrating the feasibility
\cite{China-Micius,China-Micius2,China-Austria}. 

We consider a variation of the previous setup in Fig.~\ref{fig:BB84}
applicable to satellite-based QKD experiments. 
We have the set of detectors on the right-hand side and the photon source
located on a satellite, whereas the set of detectors on the left-hand side are
located on Earth.
Within the satellite, the environment can be stabilized so that the
efficiencies of the detectors in the satellite relative to one another are
stable over time; the same is true for those located in the Earth
laboratory. These efficiency ratios can be precalibrated and recorded ahead
of time. What cannot be controlled and can vary to a large degree over time
is the relative efficiencies between the detectors on Earth and those in
space.
In particular, the photons have to pass through the atmosphere before reaching
the Earth-bound detectors.
The atmosphere acts as an absorber which only allows a rather small fraction of the
photons to pass through while leaving the polarization of the photons
essentially unchanged, and we can consider this absorption as part of the loss
in efficiency of the Earth-based detectors. 
The fraction $T$ of the photons which passes through the atmosphere depends on a
lot of factors such as temperature, humidity, atmospheric turbulence,
etc.~\cite{space2,space3,space4}.
Thus, it is hard to determine the true value of $T$; it also keeps changing
over time.
When applying the self-calibration procedure, however, the value of $T$ does
not need to be determined in advance.
Instead, it is treated as an unknown parameter which can be determined from
the data.
In this scenario, the overall efficiencies of the set of detectors on Earth
depend both on their imperfection and on losses in the atmosphere, specified
by two parameters $\eta_\textrm{right}$ and $T\eta_\textrm{left}$, to be
estimated in a self-calibration experiment. 

In the previous sections, we have assumed that the mean number of photons
$\nu$ is known.
This assumption is unrealistic in the current scenario as the power of 
the photon source might deteriorate over time and it might be hard to
calibrate it since it is located on the satellite.
Thus, we get rid of this assumption in this scenario and treat $\nu$ as
another unknown to be determined.
The likelihood for this scenario is
\begin{equation}\label{eq:likelihood2}
  L(D|\rho,\bm{\eta},\nu) = \prod_{k=1}^{24}(\nu p_k)^{n_k}\Exp{-\nu p_k}\,,
\end{equation}
up to an overall factor of no consequence (see the Appendix).
The unknowns to be determined are the eight state parameters, the mean number
of photons $\nu$, and the two efficiencies, $\eta_\text{right}$ and
$T\eta_\text{left}$. 
As $T$ is a small number, the number of detections on the satellite side is
much much greater than that on the Earth side, which makes this problem highly
asymmetric.

\begin{figure}[t]
\centering
\includegraphics[width=0.95\columnwidth]{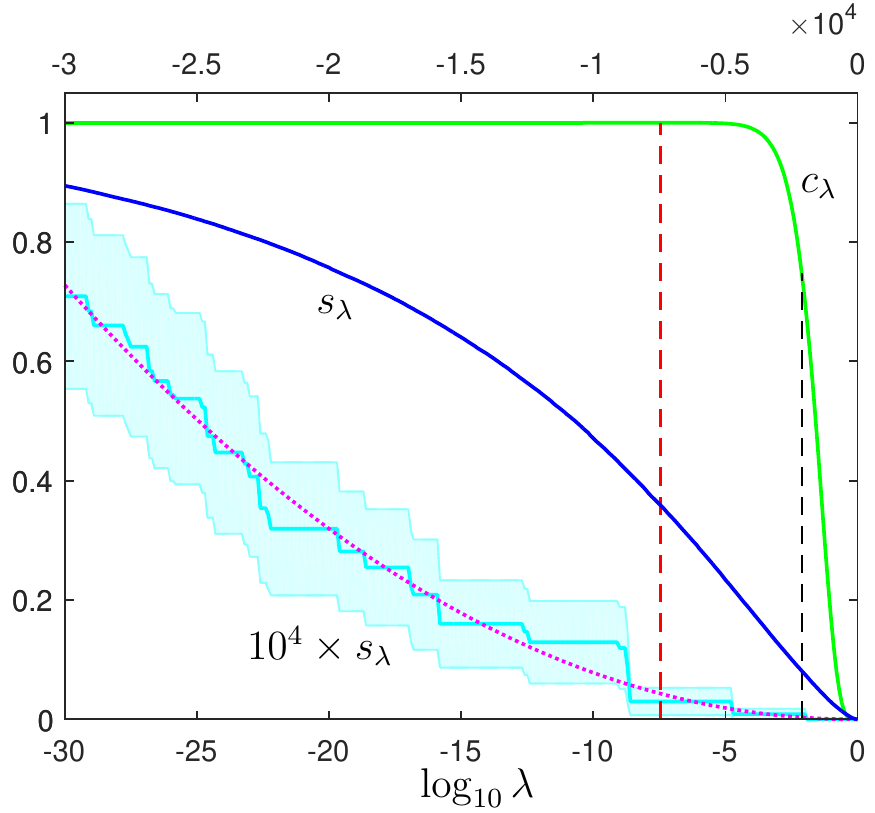}
\caption{\label{fig:extremes&c}
Size $s_\lambda$ and credibility $c_\lambda$ of the
11-dimensional OERs as a function of $\log_{10}\lambda$.
The abscissa tics at the bottom, for ${-30<\log_{10}\lambda<0}$, are for the
credibility $c_\lambda$ (green) and for the blown-up size
$10^4\times s_\lambda$ (cyan); 
the tics at the top, for ${-3\times10^4<\log_{10}\lambda<0}$,
are for the size $s_\lambda$ (blue).
The red vertical dashed line marks the critical $\lambda$ value at
$\lambda_\text{crit}=3.53\times 10^{-8}$, which identifies the plausible region.
The black vertical dashed line marks the $\lambda$ value for the mock-true state,
efficiencies, and $\nu$ value at $\lambda_\text{true}=7.73\times 10^{-3}$.
Since there are few sample points in the OERs
with ${\log_{10}\lambda>-30}$,
the size $s_{\lambda}$ is discontinuous in this range (irregular stairs in
cyan). The dotted purple line is a smooth approximation, and the band in light
cyan indicates the sampling error. 
While the precise size of the plausible region is not determined here, it is
clear that it is very small.}
\end{figure}

We obtained the following data
${D=\{}$0, 1, 0, 1, 2, 1, 2, 0, 0, 0, 1, 0, 0, 4, 5, 2, 2, 0, 3, 3, 65188,
70928, 37230, 127525$\}$ from a simulated experiment. In this example, we
assume that we do not know anything about the mock-true state, but we know
$\eta_\text{right}$, $T\eta_\text{left}$, and $\nu$ to a certain extent before
the data are taken.
Thus, we choose a uniform prior for the eight state parameters and a beta
prior for both $\eta_\text{right}$ and $T\eta_\text{left}$, 
\begin{eqnarray}
w_0(\eta_\text{right}) &\propto& (\eta_\text{right})^{55}(1-\eta_\text{right})^{15}\,,
\nonumber\\
w_0(T\eta_\text{left}) &\propto&
(T\eta_\text{left})^{\frac{1}{2}}(1-T\eta_\text{left})^{8000}\,,
\end{eqnarray}
and a gamma prior for $\nu$ ,
\begin{equation}
w_0(\nu) \propto
\nu^{99}\Exp{-\frac{\nu}{5000}}\,.
\end{equation}
For these choices, the expected values of $\eta_\text{right}$,
$T\eta_\text{left}$, and $\nu$ are $0.778$, $1.87\times10^{-4}$,
and $500\,000$ with standard deviations of $0.049$, $1.53\times10^{-4}$,
and $50\,000$, respectively.
The shortest intervals where these values lie with 0.95 prior probabilities
are $[0.681,0.870]$, $[1.97\times10^{-7},4.88\times10^{-4}]$, and
$[403716,599105]$ respectively.
This quantifies our prior guesses about these parameters and our trust in the
guessed values. 
The true values of $\eta_\text{right}$, $T\eta_\text{left}$, and $\nu$ used in the
simulation are $0.724$, $7.38\times10^{-5}$, and $500\,000$, respectively.

\begin{figure}[t]
\centering
\includegraphics[width=0.95\columnwidth]{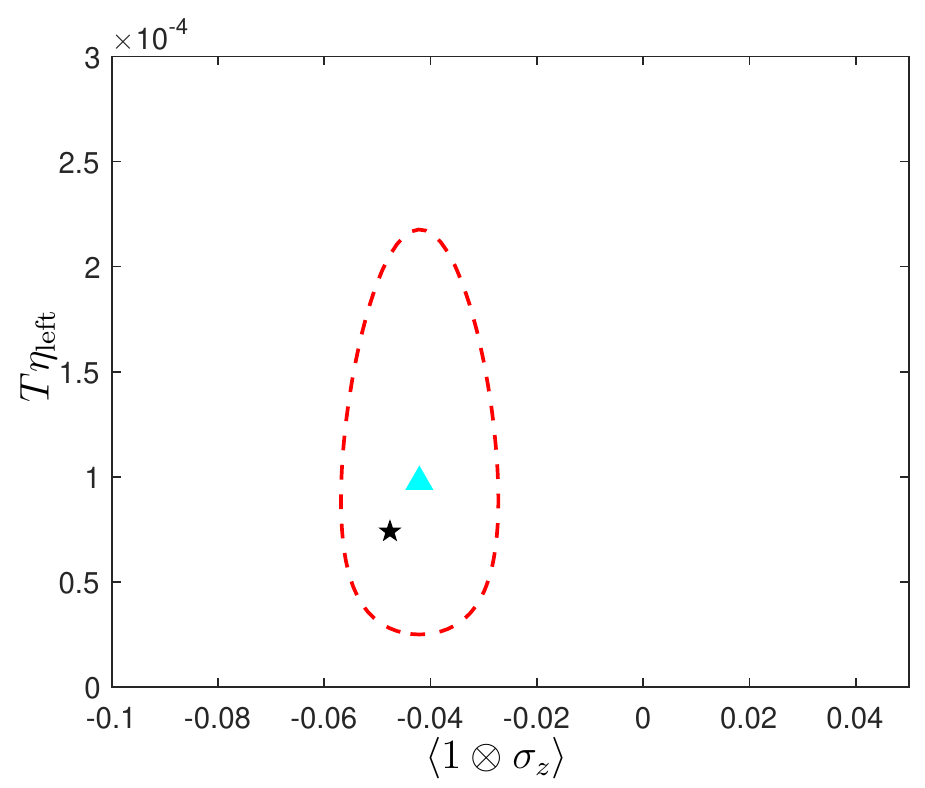}
\caption{\label{fig:2dRegion2}%
The dashed red contour shows the boundary of the two-dimensional slice of the
plausible region. 
The other parameters are set to their mock-true values.
The black star and the cyan triangle mark the mock-true values and ML
estimates, respectively.}
\end{figure}

Figure~\ref{fig:extremes&c} shows the plot for the size and credibility of the
OERs as a function of $\log_{10}\lambda$, from a Monte Carlo integration that
uses a random sample with $500\,000$ points. 
The size and credibility of the plausible region are $s\simeq4.43\times10^{-6}$ and
$c\lesssim1$, respectively; there are so few sample points in the very small
plausible region that its size cannot be determined accurately here (the
sampling error is indicated in the figure \cite{noteSamplingError}). 
The mock-true state and efficiencies are contained in the OERs with
$\lambda<7.73\times 10^{-3}$ and $c>0.7477$.
Thus, they are inside the plausible region.
The sizes of the OERs shown in Fig.~\ref{fig:extremes&c}
decrease much more rapidly than those in Fig.~\ref{fig:normals&c} because the
likelihood function is very sharply peaked in certain directions.
This is due to the large number of detections on the satellite side.
Therefore, the plausible region is a very small region around the maximum
likelihood estimator. 
Figure~\ref{fig:2dRegion2} shows the two-dimensional slice of the
11-dimensional plausible region with the other parameters set to their
mock-true values.
Again, one observes that such joint device-state error regions are much more
informative and representative of the joint error than reporting separate
error bars.

\begin{table}[t]
\centering
\caption{Mock-true and ML-estimated values for the 11 parameters
  used for the simulated data of Fig.~\ref{fig:extremes&c}.
  There are 300$\,$898 detection events in this simulation and, owing to their
  large count, some of the ML estimates of the parameters are comparable with
  their mock-true values.} 
\label{tab:mle2}
\begin{tabular}{c @{\qquad} c @{\qquad} c}
Parameter & Mock-true value & ML-estimated value   \\ \hline
$\expect{1\otimes\sigma_x}$ & -0.4062 & -0.4095 \\
$\expect{1\otimes\sigma_z}$ & -0.0478 & -0.0421 \\
$\expect{\sigma_x\otimes 1}$ & -0.1985 & -0.3878 \\
$\expect{\sigma_z\otimes 1}$ & \ 0.0085 & -0.0909  \\
$\expect{\sigma_x\otimes \sigma_x}$ & \ 0.3595 & \ 0.3190 \\
$\expect{\sigma_x\otimes \sigma_z}$ & -0.0607 & -0.1802 \\ 
$\expect{\sigma_z\otimes\sigma_x}$ & -0.0664 & -0.1180 \\
$\expect{\sigma_z\otimes\sigma_z}$ & \ 0.4192 & -0.1524 \\
$T\eta_\text{left}$ & $7.3771\times10^{-5}$ & $9.7099\times10^{-5}$ \\
$\eta_\text{right}$ & \ 0.7238 & \ 0.7435 \\
$\nu$ & 500\,000 & 486\,868
\end{tabular}
\end{table}

The ratios of detector efficiencies in this simulation are 
\begin{equation}
\eta'_{1}:\eta'_{2}:\eta'_{3}:\eta'_{4} = 0.7064 : 0.5242 : 1 : 0.3419\,,
\end{equation}
and
\begin{equation}
\eta_{1}:\eta_{2}:\eta_{3}:\eta_{4} = 0.7518 : 0.7520 : 0.6969 : 1\,.
\end{equation}
The parameters of this simulation are reported in Table~\ref{tab:mle2}.
Owing to the large count of detection events on the satellite side, some of
the estimated parameter values are quite close to their mock-true values,
while others are not. 
More importantly, however, the ML estimates are inside the plausible region,
which is of very small size.

\section{Conclusions}\label{sec:conc}
Self-calibration is needed whenever precalibration of the experimental
devices is not feasible.
The data from the self-calibration experiment yield information about both
the state as well as the measurement apparatus.
One estimates state and device parameters from the same data.
In this paper, we explain how to do this jointly, treating the state and device
parameters on equal footing.
In particular, we construct state-device optimal error regions, properly
quantifying their joint uncertainty.
We applied our methods to the example of the crosshair measurement in BB84 QKD
experiments, and discussed the case of satellite-based QKD, where
self-calibration is a necessity due to the physical situation. 

Our methods can, of course, be applied to many other situations that involve
parameter estimation. 
Examples include the absolute calibration of photon detectors in Klyshko's
scheme \cite{Klyshko80,RarityRidleyTapster87} (with or without the
simultaneous calibration of the photon-pair source), the determination of an
interferometer phase \cite{QuPhase1,QuPhase2}, and the precise estimation of
the down-conversion efficiency in experiments with entangled photons
\cite{arXiv:1808.06863}.

\acknowledgments
This work is funded by the Singapore Ministry of Education (partly through 
Tier-2 Grant No. MOE2016-T2-1-130) and the National Research Foundation of
Singapore.  
J.S. acknowledges support by the Beijing Institute of Technology Research Fund
Program for Young Scholars, the National Natural Science Foundation of China
through Grant No.~11805010, the European Research Council (Consolidator 
Grant No. 683107/TempoQ), and the Deutsche Forschungsgemeinschaft.

\section*{Appendix: Derivation of the likelihoood functions %
  in Eqs.~(\ref{eq:likelihood}) and ~(\ref{eq:likelihood2})}\label{sec:likeli}
In the lapse of time, during which we detect $N=n_1+n_2+\cdots+n_{24}$ events,
there are also $n_0$ double-null events that are not recorded, and we do not
know when they occur.
For a given sequence of $N$ detected events, interspersed by $n_0$ unrecorded
events, there are $\frac{(N+n_0)!}{N!\,n_0!}$ possible sequences of ${N+n_0}$ events.
Then, the likelihood for observing the actual sequence $S$ of detected events,
given the quantum state $\rho$, the detection efficiencies $\bm{\eta}$, and
the count $n_0$ of unrecorded double-null events, is
\begin{equation}
  L(S|\rho,\bm{\eta},n_0)
  =\frac{(N+n_0)!}{N!\,n_0!}\prod_{k=0}^{24}p_k^{n_k}\,,
\end{equation}
with the $p_k$s related to $\rho$ and $\bm{\eta}$ through Born's rule, as in
Eq.~(\ref{eq:A2}). 

The number ${N+n_0}$ of down-converted photon pairs follows a Poissonian statistic
with an average number of $\nu$ pairs during the period of data taking.
Accordingly, the likelihood for $S$, conditioned on $\rho$, $\bm{\eta}$, and $\nu$, is
\begin{eqnarray}
  L(S|\rho,\bm{\eta},\nu)
  &=&\sum_{n_0=0}^{\infty}\frac{\nu^{N+n_0}\Exp{-\nu}}{(N+n_0)!}
      L(S|\rho,\bm{\eta},n_0)
\nonumber\\ &=&\frac{\nu^N}{N!}\Exp{-(1-p_0)\nu}\prod_{k=1}^{24}p_k^{n_k}\,.
\end{eqnarray}
The actual sequence $S$ does not matter here, as only the event counts
$D=(n_1,n_2,\dots,n_{24})$ enter.
The likelihood $L(D|\rho,\bm{\eta},\nu)$ for this minimal statistic
differs from $L(S|\rho,\bm{\eta},\nu)$ by the combinatorial factor
$\frac{N!}{n_1!\,n_2!\,\cdots\,n_{24}!}$, so that
\begin{eqnarray}
  L(D|\rho,\bm{\eta},\nu)
  &=&\nu^N\Exp{-(1-p_0)\nu}\prod_{k=1}^{24}\frac{p_k^{n_k}}{n_k!}
\nonumber\\&=&\prod_{k=1}^{24}\frac{(\nu p_k)^{n_k}}{n_k!}\Exp{-\nu p_k}\,.
\end{eqnarray}
This is the product of 24 independent Poisson distributions, one for each
kind of detection event, with the average count $\nu p_k$ for the event of the
$k$th kind.
After removing the $p_k$-independent factors, which cancel in
Eq.~(\ref{eq:c-lambda}) and therefore have no bearing on the error regions of
Sec.~\ref{sec:region}, we arrive at Eq.~(\ref{eq:likelihood}). 
For the scenario in Sec.~\ref{sec:qkd}, where $\nu$ is also treated as unknown,
we have Eq.~(\ref{eq:likelihood2}) after removing only the factorial factors.

%\vfill
%%%%%%


\begin{thebibliography}{00}

\bibitem{LNP649}
\textit{Quantum State Estimation}, edited by M.~Paris and J.~\v{R}eh\'a\v{c}ek,
Lecture Notes in Physics Vol.~649 (Springer, Heidelberg, 2004).

\bibitem{Teo:16}
Y.~S. Teo,
\textit{Introduction to quantum-state estimation\/}
(World Scientific, Singapore, 2016).

\bibitem{MLEreview}
Z.~Hradil, J.~\v{R}eh\'a\v{c}ek, J.~Fiur\'a\v{s}ek, and M.~Je\v{z}ek,
\textit{Maximum-Likelihood Methods in Quantum Mechanics},
Chapter~3 in Ref.~\cite{LNP649}.

\bibitem{PRA79.020101}
D.~Mogilevtsev, J.~\v{R}eh\'a\v{c}ek, and Z.~Hradil,
\textit{Relative tomography of an unknown quantum state},
\pra~\textbf{79}, 020101(R) (2009).

\bibitem{PRA82.021807}
D.~Mogilevtsev,
\textit{Calibration of single-photon detectors using quantum statistics},
\pra~\textbf{82}, 021807 (2010).

\bibitem{NJP14.085003}
A.~M.~Bra\'{n}czyk, D.~H.~Mahler, L.~A.~Rozema, A.~Darabi,
A.~M.~Steinberg, and D.~F.~V.~James,
\textit{Self-calibrating quantum state tomography},
New~J.~Phys.\ \textbf{14}, 085003 (2012).

\bibitem{PRA87.062118}
N.~Quesada, A.~M.~Bra\'{n}czyk, and D.~F.~V.~James,
\textit{Self-calibrating tomography for multidimensional systems},
\pra~\textbf{87}, 062118 (2013).

\bibitem{njp19.043003}
B.~P. Williams and P. Lougovski,
\textit{Quantum state estimation when qubits are lost: %
  a no-data-left-behind approach},
New J. Phys.\ \textbf{19}, 043003 (2017).

\bibitem{OER13}
J.~Shang, H.~K.~Ng, A.~Sehrawat, X.~Li, and B.-G.~Englert,
\textit{Optimal error regions for quantum state estimation},
New~J.~Phys.\ \textbf{15}, 123026 (2013).

\bibitem{BB84}
C.~H. Bennett and G. Brassard, 
\textit{Quantum cryptography: Public key distribution and coin tossing},
in \emph{Proceedings of the IEEE International Conference on Computers, Systems and Signal Processing}, 
(IEEE, New York, 1984), Vol. 175.

\bibitem{Evans+2:06}
M.~J. Evans, I. Guttman, and T. Swartz,
\textit{Optimality and computations for relative surprise inferences.},
Can.~J.~Stat.\ \textbf{34}, 113 (2006).

\bibitem{Evans:15}
M. Evans, \textit{Measuring Statistical Evidence Using Relative Belief\/},
Monographs on Statistics and Applied Probability, Vol.~144
(CRC Press, Boca Raton, 2015).

\bibitem{OEI16}
X. Li, J. Shang, H.~K. Ng, and B.-G. Englert,
\textit{Optimal error intervals for properties of the quantum state},
\pra~\textbf{94}, 062112 (2016).

\bibitem{Evans-essay-2019}
M. Evans, \textit{The Measurement of Statistical Evidence as the Basis for
  Statistical Reasoning},
eprint arXiv:1906.09484 [math.ST] (2019).

\bibitem{Evans-Guo-2019}
M. Evans and Y. Guo, \textit{Measuring and Controlling Bias for some Bayesian
  Inferences and Relation to Frequentist Criteria},
eprint arXiv:1903.01696 [math.ST] (2019).

\bibitem{Shang+1:15}
J.~Shang, Y.-L.~Seah, H.~K.~Ng, D.~J.~Nott, and B.-G. Englert,
\textit{Monte Carlo sampling from the quantum state space.~I},
New~J.~Phys. \textbf{17}, 043017 (2015).

\bibitem{Seah+1:15}
Y.-L.~Seah, J.~Shang, H.~K.~Ng, D.~J.~Nott, and B.-G. Englert,
\textit{Monte Carlo sampling from the quantum state space.~II},
New~J.~Phys. \textbf{17}, 043018 (2015).

\bibitem{YS19}
C.~Oh, Y.~S.~Teo, and H.~Jeong,
\textit{Efficient Bayesian credible-region certification for quantum-state
  tomography}, 
\pra~\textbf{100}, 012345 (2019).

\bibitem{space}
R. Ursin, T. Jennewein, J. Kofler, J. M. Perdigues, L. Cacciapuoti,
C. J. de Matos, M. Aspelmeyer, A. Valencia, T. Scheidl, A. Acin, C. Barbieri,
G. Bianco, C. Brukner, J. Capmany, S. Cova, D. Giggenbach, W. Leeb,
R. H. Hadfield, R. Laflamme, N. L\"utkenhaus, G. Milburn, M. Peev, T. Ralph,
J. Rarity, R. Renner, E. Samain, N. Solomos, W. Tittel, J. P. Torres,
M. Toyoshima, A. Ortigosa-Blanch, V. Pruneri, P. Villoresi, I. Walmsley,
G. Weihs, H. Weinfurter, M. Zukowski, and A. Zeilinger,
\textit{Space-quest, experiments with quantum entanglement in space},
Europhysics News \textbf{40}, 26 (2009).
  
\bibitem{China-Micius}
S.-K. Liao, W.-Q. Cai, W.-Y. Liu, L. Zhang, Y. Li, J.-G. Ren, J. Yin, Q. Shen,
Y. Cao, Z.-P. Li, F.-Z. Li, X.-W. Chen, L.-H. Sun, J.-J. Jia, J.-C. Wu,
X.-J. Jiang, J.-F. Wang, Y.-M. Huang, Q. Wang, Y.-L. Zhou, L. Deng, T. Xi,
L. Ma, T. Hu, Q. Zhang, Y.-A. Chen, N.-L. Liu, X.-B. Wang, Z.-C. Zhu,
C.-Y. Lu, R. Shu, C.-Z. Peng, J.-Y. Wang, and J.-W. Pan,  
\textit{Satellite-to-ground quantum key distribution},
Nature \textbf{549}, 43 (2017).   
  
\bibitem{China-Micius2}
J. Yin, Y. Cao, Y.-H. Li, J.-G. Ren, S.-K. Liao, L. Zhang, W.-Q. Cai,
W.-Y. Liu, B. Li, H. Dai, M. Li, Y.-M. Huang, L. Deng, L. Li, Q. Zhang,
N.-L. Liu, Y.-A. Chen, C.-Y. Lu, R. Shu, C.-Z. Peng, J.-Y. Wang,
and J.-W. Pan,
\textit{Satellite-to-ground entanglement-based quantum key distribution},
\prl \textbf{119}, 200501 (2017).  
  
\bibitem{China-Austria}
S.-K. Liao, W.-Q. Cai, J. Handsteiner, B. Liu, J. Yin, L. Zhang, D. Rauch,
M. Fink, J.-G. Ren, W.-Y. Liu, Y. Li, Q. Shen, Y. Cao, F.-Z. Li, J.-F. Wang,
Y.-M. Huang, L. Deng, T. Xi, L. Ma, T. Hu, L. Li, N.-L. Liu, F. Koidl,
P. Wang, Y.-A. Chen, X.-B.Wang, M. Steindorfer, G. Kirchner, C.-Y. Lu,
R. Shu, R. Ursin, T. Scheidl, C.-Z. Peng, J.-Y. Wang, A. Zeilinger,
and J.-W. Pan,
\textit{Satellite-relayed intercontinental quantum network},
\prl \textbf{120}, 030501 (2018).

\bibitem{space2}
R. Ursin, F. Tiefenbacher, T. Schmitt-Manderbach, H. Weier, T. Scheidl,
M. Lindenthal, B. Blauensteiner, T. Jennewein, J. Perdigues, P. Trojek,
B. \"Omer, M. F\"urst, M. Meyenburg, J. Rarity, Z. Sodnik, C. Barbieri,
H. Weinfurter, and A. Zeilinger,
\textit{Entanglement-based quantum communication over 144\,km},
Nature Phys.\ \textbf{3}, 481 (2007).

\bibitem{space3}
R. Bedington, J.~M. Arrazola, and A. Ling, 
\textit{Progress in satellite quantum key distribution},
npj Quantum Inf.\ \textbf{3}, 30 (2017).

\bibitem{space4}
D.~K.~L. Oi, A. Ling, G. Vallone, P. Villoresi, S. Greenland, E. Kerr,
M. Macdonald, H. Weinfurter, H. Kuiper, E. Charbon, and R. Ursin,
\textit{CubeSat quantum communications mission},
EPJ Quantum Technology \textbf{4}, 6 (2017).

\bibitem{noteSamplingError}
The sampling error can be determined as in Sec.~VI A2 in
Ref.~\cite{arXiv:1808.06863}. 

\bibitem{Klyshko80}
D. N. Klyshko,
\textit{Use of two-photon light for absolute calibration of photo-electric
    detectors}, 
Sov.\ J.\ Quant.\ Electron.\ \textbf{10}, 1112 (1980).

\bibitem{RarityRidleyTapster87}
J. D. Rarity,  K. D.  Ridley, and P. R. Tapster,
\textit{Absolute  measurement  of  detector  quantum  efficiency  
  using parametric downconversion}, 
Appl.\ Opt.\ \textbf{26}, 4616 (1987).

\bibitem{QuPhase1}
Z. Hradil, R. My\v{s}ka, J. Pe\v{r}ina, M. Zawisky, Y. Hasegawa, and H. Rauch,
\textit{Quantum Phase in Interferometry},
\prl \textbf{76}, 4295 (1996).

\bibitem{QuPhase2}
J. \v{R}eh\'a\v{c}ek, Z. Hradil, M. Zawisky, S. Pascazio, H. Rauch, and
J. Pe\v{r}ina,
\textit{Testing of quantum phase in matter-wave optics},
\pra \textbf{60}, 473 (1999).

\bibitem{arXiv:1808.06863}
Y. Gu, W. Li, M. Evans, and B.-G. Englert,
\textit{Very strong evidence in favor of quantum mechanics and against local
  hidden variables from a Bayesian analysis}, 
\pra\textbf{99}, 022112 (2019).

\end{thebibliography}
\end{document}